\documentclass[12pt]{article}

\usepackage[margin=1in]{geometry}
\usepackage{setspace}
\usepackage{amsmath, amssymb, amsthm}
\usepackage{graphicx}
\usepackage{booktabs}
\usepackage{threeparttable}
\usepackage{multirow}
\usepackage{array}
\usepackage{caption}
\usepackage{subcaption}
\usepackage{siunitx}
\usepackage{hyperref}
\usepackage[nameinlink,noabbrev]{cleveref}
\usepackage{natbib}

\hypersetup{
  colorlinks=true,
  linkcolor=blue,
  citecolor=blue,
  urlcolor=blue
}

\newcommand{\OAIRate}{\textit{OAI rate}}

\newcommand{\OAI}{\textit{OAI}}

\usepackage{tabularx}
 \newcolumntype{L}[1]{>{\raggedright\arraybackslash}p{#1}}
\newcolumntype{C}[1]{>{\centering\arraybackslash}p{#1}}
\newcolumntype{R}[1]{>{\raggedleft\arraybackslash}p{#1}}

\title{\vspace{-1.5em} Scale and Capacity Limits in Decentralized FDA Food-Safety Enforcement}
\author{Guy Tchuente\footnote{Department of Agricultural Economics, e-mail: \texttt{gtchuent@purdue.edu}}\\
  Purdue University}
\date{\today}

\begin{document}
\maketitle

\begin{abstract}
\noindent

This paper asks whether regulatory monitoring exhibits nonlinear capacity limits as the scale and complexity of the regulated environment increase. Using a county--year panel of U.S. Food and Drug Administration (FDA) inspections merged with local establishment counts, we identify a sharp breakpoint: beyond a threshold scale, severe inspection findings rise while inspection effort per establishment flattens or declines. The threshold and the post-break deterioration vary across food-related industry groups and shift with proxies for local density and connectedness, consistent with monitoring becoming ``too big to monitor" in more interconnected production environments rather than driven by simple reallocation or delay. Methodologically, we provide a portable breakpoint-selection and piecewise-estimation framework that can be applied to other enforcement settings.

\end{abstract}

\vspace{0.5em}

\newpage

\section{Introduction}

The resources required to monitor compliance often grow faster than the size of the regulated environment.\footnote{Regulatory monitoring has first-order welfare implications because it disciplines behavior in settings with large externalities. In food and drug markets, failures of oversight can translate into preventable health risks, costly recalls, and persistent noncompliance that spills across supply chains.} When inspection capacity, follow-up, and administrative bandwidth do not expand proportionally with the number of regulated establishments, and when compliance risks are correlated through local supply chains and shared infrastructure, monitoring performance may deteriorate \emph{nonlinearly}.

This paper studies whether such ``too big to monitor'' dynamics arise in U.S.\ Food and Drug
Administration (FDA) inspections, and whether the scale at which monitoring breaks down depends on
local density and connectedness.

We assemble a county--year panel of FDA inspections over 2009--2019 and measure monitoring outcomes
using inspection classifications. Our primary outcome is the probability that an inspection results in an Official Action Indicated (OAI) finding, a marker of serious observed noncompliance.\footnote{FDA inspections are classified into {NAI/VAI/OAI}; an {OAI} is the most severe inspection classification and indicates that the agency views regulatory action as warranted (e.g., enforcement follow-up and corrective actions). In economics and closely related empirical work using FDA inspection microdata, OAI is therefore treated as a high-salience, policy-relevant measure of
serious compliance failures and regulatory intervention intensity. See \citet{galdin2024resilience,macher2011regulator}
for a discussion of FDA inspection outcomes and their interpretation, and \citet{wang2025oai}
for econometric evidence linking OAI outcomes to subsequent supply-chain outcomes (drug shortages).} We also measure monitoring
effort per establishment using County Business Patterns (CBP) establishment counts as a proxy for
the local regulated population. The baseline specification includes county economic controls
(poverty, income, unemployment) and a rich set of fixed effects.

Our empirical approach is intentionally simple and portable. We model outcomes as a piecewise-linear
function of local scale (in logs in the baseline), allowing both a discontinuity (a ``jump'') and a
change in slope (a ``kink'') at an unknown cutoff $c^{\star}$. We select $\widehat c^{\star}$ by a
grid search that maximizes the grouped binomial-logit likelihood for OAI counts, and we examine
alternative objectives and candidate grids in robustness checks. After selecting the cutoff, we
estimate jump and kink effects with the same fixed effects and controls, clustering standard errors
by county. Conceptually, the cutoff captures \emph{where} monitoring begins to break down, while the
post-cutoff slope captures \emph{how quickly} performance changes as local scale increases further.\\

\paragraph{Main empirical findings.}
We establish three results.
\begin{enumerate}
  \item \textbf{A national-scale breakpoint.} In pooled county--year data, monitoring outcomes exhibit
  a clear regime change at a threshold $\widehat c^{\star}$. Above this cutoff, adverse outcomes rise
  more steeply with scale (a positive kink in OAI-type measures), while monitoring effort per
  establishment flattens or declines (a negative kink in effort-type measures). These qualitative
  patterns persist across alternative specifications.
  \item \textbf{Heterogeneity in where breakdown occurs.} Allowing thresholds to vary by industry group
  reveals substantial dispersion in $\widehat c^{\star}_g$, consistent with monitoring limits that are
  \emph{local} and \emph{sector-specific}, rather than governed by a single universal county-size
  threshold.
  \item \textbf{Density and connectedness matter.} Within industry groups, estimated thresholds and
  post-cutoff slope changes shift systematically across within-group density bins. This pattern is
  consistent with the view that what matters is not only the number of establishments, but also the
  organization of production: in denser and more interconnected environments, correlated risks and
  more interdependent follow-up can make effective monitoring deteriorate more abruptly.
\end{enumerate}

These findings are consistent with a ``too big to monitor'' mechanism in which monitoring technology
changes once local scale exceeds effective capacity. Standard enforcement models emphasize that
detection is costly and enforcement resources are limited, so inspection intensity need not scale
one-for-one with the number of regulated entities
\citep{becker1968crime,polinsky2000public,harrington1988enforcement}. In the FDA setting, staffing,
travel time, administrative processing, and enforcement follow-up may not keep pace with local scale,
putting downward pressure on inspections per establishment beyond $\widehat c^{\star}$. At the same
time, a complementary interpretation emphasizes \emph{spillovers and connectedness}: if violations are
correlated through local production networks, diagnosing and remediating problems can require more
coordination per case, so effective coverage can fall even when nominal inspection activity does not
decline one-for-one \citep{carvalho2014networks,acemoglu2015systemic}. Our heterogeneity results---and
auxiliary evidence that provides limited support for simple triage or queueing-delay explanations---are
most consistent with connectedness and local complexity playing a central role in the breakdown at scale.

A central concern in threshold analyses is that algorithms may find ``breaks'' mechanically. We
therefore emphasize two safeguards. First, the qualitative post-cutoff patterns survive across
alternative scale definitions, objective functions, candidate grids, and fixed-effect menus. Second,
placebo and balance checks using predetermined county characteristics (poverty, income, unemployment)
show no meaningful jump or kink at the baseline cutoff, and applying the same cutoff-search procedure
to these predetermined outcomes does not produce a stable ``false threshold.'' Together, these
exercises support an interpretation in which the estimated breakpoint reflects a monitoring-relevant
regime change rather than a generic nonlinear trend.

The paper contributes to three literatures. Substantively, it provides evidence that regulatory
monitoring can break down nonlinearly with local scale in a core enforcement setting. Empirically, it
documents that the scale at which breakdown occurs varies across sectors and shifts with within-group
density, supporting a ``too big to monitor locally'' interpretation. Methodologically, it offers a
replicable framework for breakpoint detection and piecewise estimation in panel data that can be
transported to other oversight contexts (e.g., workplace safety, environmental inspections),
complementing work on thresholds and kink-type nonlinearities
\citep{hansen1999threshold,card2015rkd}.

Section~\ref{sec:background} describes FDA inspections and outcome construction.
Section~\ref{sec:data} introduces the county--year panel and scale and density proxies.
Section~\ref{sec:design} outlines breakpoint detection and estimation.
Section~\ref{sec:results} presents the main results, heterogeneity by group and density, and the
mechanism discussion together with robustness and placebo checks.
Section~\ref{sec:conclusion} concludes.

\section{Institutional Background and Measurement}\label{sec:background}

\subsection{Regulatory split and enforcement architecture}

Food safety oversight in the United States is divided across agencies with
distinct enforcement models. For the segments we study, the \emph{Food and Drug
Administration} (FDA) relies on a \emph{decentralized inspection and enforcement
architecture}: inspections are executed through a geographically distributed
field organization (within the Office of Regulatory Affairs) and, critically,
through formal partnerships with state agencies that conduct inspections under
FDA authority and procedures.\footnote{Throughout, we use ``decentralized'' in
an administrative sense: operational capacity is implemented via multiple field
units and contracted state partners, rather than a single nationally deployed
in-plant inspection corps.} \citep{FDA_ORA_Directory, FDA_CP7303_040}

This model contrasts with the \emph{U.S. Department of Agriculture}'s Food Safety
and Inspection Service (FSIS), which administers federal inspection in
federally inspected slaughter and processing plants via a large inspection
workforce operating under a centralized field-operations chain of command.
\citep{USDAOIG_FSIS_Inspection_2013}

\subsection{FDA inspections and the state contracting channel}\label{subsec:fda_process}

FDA conducts inspections of facilities under its jurisdiction to assess
compliance with applicable requirements. A key feature for our setting is that
FDA does not rely exclusively on federal inspectors: it operates a \emph{Human
Food Contract Inspection Program} in which \emph{state contract inspections} are
a core input into routine oversight.\citep{FDA_CP7303_040} Under this program,
state agencies perform inspections of FDA-regulated firms pursuant to FDA
contracts and associated operational requirements (training/credentialing,
inspection protocols, documentation and reporting to FDA systems, and
coordination with FDA field management).\citep{FDA_CP7303_040} This contracting
channel creates meaningful scope for local capacity constraints and monitoring
trade-offs to interact with the scale of the regulated sector in a place.

Inspection outcomes are classified into categories including ``No Action
Indicated'' (NAI), ``Voluntary Action Indicated'' (VAI), and ``Official Action
Indicated'' (OAI).\citep{FDA_LifeAfterOAI} We focus on OAI as a measure of
\emph{severe} findings---outcomes that typically involve significant regulatory
deviations and trigger follow-up actions by the agency.\citep{FDA_LifeAfterOAI}

\subsection{Key outcomes and measurement}\label{subsec:measurement}

We construct outcomes at the county-year level after collapsing the underlying
inspection records and establishment counts to a unique county-year panel.

\begin{itemize}
  \item \textbf{Performance outcome:} \OAIRate, defined as
  \[
    \OAIRate_{ct} \equiv \frac{\text{OAI}_{ct}}{\text{Inspections}_{ct}}.
  \]
  \item \textbf{Effort outcome:} inspection intensity per establishment, measured as
  \[
    \text{Effort}_{ct} \equiv \ln(\text{Inspections}_{ct}) - \ln(\text{Establishments}_{ct}),
  \]
  where the establishment count is constructed from the relevant establishment
  universe for FDA-regulated food facilities in the county-year.

\end{itemize}

\subsection{Why decentralization matters at FDA}\label{subsec:why_decentralization}

\citet{tchuente2025too} develops a monitoring theory in networks in which
the oversight capacity is locally limited and the spillovers of information (or enforcement) propagate through connected units. A central prediction is the
existence of \emph{scale thresholds}: as the size or density of the monitored
network increases, monitoring can remain effective up to a point and then
\emph{break down} nonlinearly once congestion overwhelms local capacity. The
breakpoint is not universal; it shifts with parameters that govern spillovers
and network density, so that more connected environments can exhibit breakdown at different scales.

FDA oversight provides a natural setting to assess these predictions because its
enforcement system is operationally decentralized. Inspections are implemented
through a geographically distributed federal field structure and through state
contract inspections conducted under FDA authority.\citep{FDA_ORA_Directory, FDA_CP7303_040}
As a result, monitoring capacity is not a single national resource smoothly
allocated across space. Instead, effective oversight in a county depends on
(i) local federal and/or state partner capacity, (ii) the scale and composition
of the regulated sector, and (iii) the feasibility of reallocating inspection
effort across locations when congestion rises.\citep{FDA_CP7303_040} In the logic
of \citet{tchuente2025too}, this institutional structure creates precisely the
conditions under which monitoring breakdown can appear as a \emph{kink} or
\emph{jump} in performance and effort outcomes at an endogenous threshold.
Accordingly, we focus on identifying breakpoints in OAI outcomes and inspection
intensity, and on testing whether the location of these thresholds varies with
local density in ways consistent with the theory.

\section{Data and Construction of Scale and Density Measures}
\label{sec:data}

\subsection{Data sources and county mapping}

We combine four sources. First, FDA inspection outcomes come from the agency's
inspection classification records, which report inspection dates, outcomes
(including NAI/VAI/OAI), and facility identifiers and location fields (e.g.,
address/ZIP) \citep{FDAInspectionClassificationDB, FDADataDashboard}.
Second, we measure local regulated scale using the U.S.\ Census Bureau's County
Business Patterns (CBP), which provides annual establishment counts by county
and industry (NAICS) \citep{CBPGeography}.
Third, because FDA location fields are often recorded at the ZIP-code level,
we use a ZIP-to-county crosswalk to map facilities into counties in a systematic
and replicable way \citep{HUDUSPSCrosswalk}.
Fourth, we incorporate predetermined county-year covariates from Census SAIPE
(poverty rate and median household income) and BLS LAUS (unemployment rate)
to flexibly absorb local socioeconomic conditions \citep{CensusSAIPE, BLSLAUS}.

\subsection{County-year panel}

The unit of observation is county $c$ in year $t$. After mapping facilities to
counties, we aggregate FDA inspections to the county-year level. Let
$Inspections_{ct}$ denote the total number of inspections and $\OAI_{ct}$ the
number classified as OAI. We also track the number of \emph{unique} facilities
inspected in a county-year using facility identifiers (FEI) to avoid conflating
repeat inspections of the same facility with extensive margin coverage. The main
outcomes are:
\begin{itemize}
  \item \textbf{Inspection volume:} $Inspections_{ct}$.
  \item \textbf{Severe findings:} $\OAI_{ct}$.
  \item \textbf{Performance:} $\OAIRate_{ct} \equiv \OAI_{ct}/Inspections_{ct}$.
\end{itemize}
In the baseline sample, we restrict attention to county-years with positive CBP
establishments and positive inspection activity, and we apply minimum-support
requirements in robustness checks to ensure adequate observations on both sides
of candidate thresholds.

\subsection{Scale measures}

We construct measures of local regulated scale from CBP establishment counts.
Let $S_{ct}$ denote the relevant scale (running variable). In the baseline
specification, $S_{ct}$ is the total number of food-related establishments in
county $c$ and year $t$ (constructed from CBP NAICS tabulations). For
heterogeneity analyses, we also form group-specific scale measures,
$S_{g,ct}$, by summing CBP establishments within NAICS groups $g$ that match our
food-sector categories. The primary running variable in estimation is
$\ln(S_{ct})$ (or $\ln(S_{g,ct})$ in group-specific analyses).

\subsection{Connectedness proxies}

To proxy for local connectedness and the scope for spillovers in monitoring, we
use density-type measures derived from CBP establishment counts and their
composition. Our baseline proxy is a log-density measure,
\[
D_{ct} \equiv \ln(\text{FoodEstablishments}_{ct}),
\]
and in within-group analyses we form terciles (low/medium/high) of $D_{ct}$
computed within each group-specific sample. This binning strategy operationalizes
the hypothesis that thresholds may shift with local density/connectedness---that
is, with the intensity of interactions and the practical difficulty of monitoring
larger or more interconnected regulated environments \citep{tchuente2025too}.

\begin{table}[!htbp]\centering
\caption{Summary statistics (county-year, 2009--2019)}\label{tab:summarystats} \small
\begin{tabular}{lrrrrrrrr}\hline
Variable & N & Mean & SD & P25 & P50 & P75 & Min & Max\\\hline
OAI rate (OAI/inspections) & 4027 & 0.04 & 0.06 & 0.00 & 0.00 & 0.06 & 0.00 & 0.62 \\
ln(CBP establishments) & 4027 & 3.32 & 0.82 & 2.64 & 3.18 & 3.81 & 2.30 & 6.95 \\
CBP establishments & 4027 & 43.45 & 72.13 & 14.00 & 24.00 & 45.00 & 10.00 & 1045.00 \\
FDA inspections & 4027 & 35.26 & 50.52 & 11.00 & 20.00 & 40.00 & 5.00 & 848.00 \\
OAI outcomes & 4027 & 1.43 & 3.11 & 0.00 & 0.00 & 2.00 & 0.00 & 48.00 \\
Unique facilities (FEI) inspected & 4027 & 27.00 & 38.36 & 9.00 & 16.00 & 31.00 & 1.00 & 655.00 \\
Inspections per establishment & 4027 & 0.90 & 0.51 & 0.56 & 0.79 & 1.10 & 0.14 & 5.68 \\
ln(insp per est) = ln(insp)-ln(est) & 4027 & -0.24 & 0.52 & -0.59 & -0.24 & 0.09 & -1.99 & 1.74 \\
OAI events per establishment & 4027 & 0.03 & 0.05 & 0.00 & 0.00 & 0.05 & 0.00 & 0.50 \\
SAIPE poverty rate (all ages), percent & 4027 & 13.76 & 5.02 & 10.10 & 13.50 & 17.00 & 3.10 & 38.10 \\
ln(SAIPE median HH income) & 4027 & 10.94 & 0.25 & 10.76 & 10.90 & 11.10 & 10.32 & 11.93 \\
LAUS unemployment rate (annual avg, \%) & 4027 & 6.61 & 2.87 & 4.40 & 6.20 & 8.30 & 1.60 & 29.10 \\
\hline\end{tabular}
\end{table}

\section{Empirical Strategy: Detecting Monitoring Congestion}\label{sec:design}

We test the central prediction in \citet{tchuente2025too} that monitoring
performance can deteriorate nonlinearly once the local scale of regulated
activity becomes sufficiently large. Our empirical strategy has two parts:
(i) a piecewise specification that allows a discrete change and a slope change
at a candidate threshold, and (ii) a grid-search procedure that selects the
threshold that best fits the data in a pre-specified search window.

\subsection{Piecewise threshold specification}

Let $S_{ct}$ denote the county-year scale measure (Section~\ref{sec:data}) and
define $x_{ct}\equiv \ln(S_{ct})$. For a candidate cutoff $c>0$, define
\[
\mathbf{1}_{ct}(c)\equiv \mathbf{1}\{S_{ct}>c\},
\qquad
\mathrm{After}_{ct}(c)\equiv \mathbf{1}\{S_{ct}>c\}\big(x_{ct}-\ln(c)\big).
\]
We estimate the kinked specification
\begin{equation}
y_{ct}
= \alpha + \beta x_{ct}
+ \gamma\, \mathbf{1}_{ct}(c)
+ \delta\, \mathrm{After}_{ct}(c)
+ \lambda_t + \mu_s + Z'_{ct}\theta + \varepsilon_{ct},
\label{eq:piecewise}
\end{equation}
where $\lambda_t$ and $\mu_s$ are year and state fixed effects, and $Z_{ct}$
includes county controls (SAIPE poverty rate, $\ln$ median household income,
and LAUS unemployment rate, as available). The coefficient $\gamma$ captures a
level shift at the cutoff (``jump''), while $\delta$ captures the change in the
marginal relationship between the outcome and log scale above the cutoff
(``kink''). Standard errors are clustered at the county level.

\paragraph{Outcomes and weights.}
We implement \eqref{eq:piecewise} for (i) monitoring performance and (ii)
monitoring effort. For performance, we use the county-year OAI rate
$y_{ct}=\mathrm{OAI}_{ct}/\mathrm{Inspections}_{ct}$ and weight observations by
$\mathrm{Inspections}_{ct}$ so that high-volume county-years contribute
proportionately. For effort, we use
$y_{ct}=\ln(\mathrm{Inspections}_{ct})-\ln(\mathrm{Establishments}_{ct})$ and
weight by $\mathrm{Establishments}_{ct}$ (or the relevant denominator), which
reflects exposure.

\subsection{Selecting the cutoff by grid search}

The cutoff $c$ is not known a priori. We select it by grid search over a
candidate set $\mathcal{C}$ constructed from the empirical distribution of
$S_{ct}$ in a pre-specified \emph{search sample} (e.g., restricting to
$t\leq\overline t$). We restrict candidates to those with sufficient support on
both sides of the threshold (minimum observations below and above $c$).

Our primary objective matches the binomial inspection micro-foundation:
\[
\mathrm{OAI}_{ct}\sim \mathrm{Binomial}(\mathrm{Inspections}_{ct},p_{ct}),
\qquad
\mathrm{logit}(p_{ct})
= \alpha + \beta x_{ct}
+ \gamma\, \mathbf{1}_{ct}(c)
+ \delta\, \mathrm{After}_{ct}(c)
+ \lambda_t + \mu_s + Z'_{ct}\theta .
\]
For each $c\in\mathcal{C}$ we estimate the binomial logit and record its
log-likelihood $\ell(c)$. The selected cutoff is
\[
\widehat{c}\in \arg\max_{c\in\mathcal{C}} \ \ell(c).
\]
As a robustness check, we also consider an alternative objective that minimizes
the weighted residual sum of squares from the linear-probability version of
\eqref{eq:piecewise}.

\subsection{Estimation and inference}

Given $\widehat c$, we re-estimate \eqref{eq:piecewise} on the full analysis
sample to obtain $(\hat\beta,\hat\gamma,\hat\delta)$ for each outcome. We report
cluster-robust standard errors for the second-step estimates, clustering at the
county level.

Because $\widehat c$ is a discrete argmax over a finite grid, its sampling
distribution is nonstandard and can be sensitive to grid construction. In
principle, one can resample counties and re-run the search to obtain
uncertainty for functions of $\widehat c$. In practice, with year and state
fixed effects and county controls, subsample-specific estimation can become
fragile (the cutoff may fail to be identified in many resamples when candidate
support is thin). For this reason, our main inference focuses on the piecewise
effects conditional on the selected cutoff, and we interpret subsample cutoff
comparisons primarily through their relative ranking and robustness across
specifications.\footnote{In our application, both cluster bootstrap and
leave-one-county-out jackknife procedures often fail to identify cutoffs in a
large share of resamples once fixed effects and controls are included, so
resampling-based confidence intervals for cutoff differences are not stable.}

\paragraph{Interpretation and identification.}
Our empirical design does not rely on an externally imposed threshold or quasi-random assignment around a policy cutoff. Instead, we estimate a breakpoint $c^{\star}$ that best summarizes a nonlinear relationship between monitoring outcomes and regulated scale. Formally, $c^{\star}$ is the maximizer of a well-defined fit criterion (log-likelihood or weighted RSS) for a piecewise specification of $\mathbb{E}[Y_{ct}\mid S_{ct},\text{FE},X_{ct}]$. The resulting ``jump'' and ``kink'' terms should therefore be interpreted as reduced-form features of the conditional expectation function rather than as causal effects of crossing an exogenous threshold. To guard against data-mining concerns, we (i) document stability across alternative candidate grids, objective functions, and fixed-effect structures, and (ii) implement placebo exercises using predetermined county characteristics.

\section{Results}\label{sec:results}

\subsection{National cutoff: evidence of a monitoring ``cliff''}

Table~\ref{tab:breakpoints} reports the national breakpoint estimate and the
associated piecewise effects. Two patterns stand out. First, the OAI rate
exhibits a clear increase at the threshold: the estimated discontinuity is
positive, and the post-cutoff slope is steeper than below the cutoff. In the
context of monitoring performance, this implies that once county scale crosses
$\widehat c=71$, adverse inspection outcomes become more likely and increase
more quickly with further increases in scale.

Second, monitoring effort shows weaker evidence of an immediate discontinuity
but a negative post-cutoff slope (kink). Put differently, beyond the cutoff,
inspections per establishment grow more slowly (and may decline), even as the
OAI rate rises. This ``divergence'' between outcomes and effort is consistent
with monitoring effectiveness deteriorating at scale. Importantly, this reduced
coverage is a broad (county-level) margin; below we show limited evidence that
it is explained by simple reallocations toward repeat offenders or by systematic
delays in OAI follow-up scheduling, which instead points to mechanisms related
to the complexity of monitoring in large and dense local production environments.

\begin{table}[!htbp]\centering
\caption{Breakpoint estimates ($c^\star=71$, $\ln(c^\star)=4.2627$). Year and state fixed effects plus county controls.}
\label{tab:breakpoints}
\footnotesize
\setlength{\tabcolsep}{6pt}
\renewcommand{\arraystretch}{1.15}
\begin{tabular}{l r cc cc}
\toprule
Outcome & $N$ & Jump & Kink \\
\midrule
OAI rate ($\mathrm{OAI}/\mathrm{Inspections}$)
  & 4{,}027 & 0.0060\,(0.0665) & 0.0097\,(0.0010) \\
Effort: $\ln(\mathrm{Insp})-\ln(\mathrm{Est})$
  & 4{,}027 & 0.0541\,(0.3726) & -0.1034\,(0.0890) \\
OAI per establishment ($\mathrm{OAI}/\mathrm{Est}$)
  & 4{,}027 & 0.0049\,(0.0978) & 0.0022\,(0.4300) \\
\bottomrule
\end{tabular}

\vspace{0.4em}
\parbox{0.95\linewidth}{\footnotesize \textit{Notes:} Unit of observation is the county--year ($N$ county--years).
$c^\star$ is selected by maximizing the grouped binomial-logit likelihood for OAI counts using year and state fixed effects and county controls.
``Jump'' is the discontinuity at $\ln(c^\star)$ and ``Kink'' is the change in slope above $\ln(c^\star)$ in a piecewise-linear specification in $\ln(\mathrm{Est})$.
Parentheses report two-sided $p$-values; standard errors are clustered by county.}
\end{table}

\paragraph{Cutoff uncertainty.}
To summarize uncertainty in the selected breakpoint, we construct a likelihood--ratio
profile over the candidate grid. Let $\ell(c)$ denote the binomial-logit log-likelihood
evaluated at candidate cutoff $c$, and let $\ell_{\max}=\max_{c}\ell(c)$.
We define
\[
LR(c)=2\bigl(\ell_{\max}-\ell(c)\bigr),
\]
and use the $\chi^2_1$ reference cutoff $\chi^2_{1,\,1-\alpha}$ (with $\alpha=0.05$).
The 95\% profile set is $\{c: LR(c)\le \chi^2_{1,0.95}\}$. In our baseline specification,
this yields a 95\% profile confidence interval of $[68,95]$ for $c^\star$.

\subsection{RDD-style visualization around the cutoff}

\begin{figure}[!htbp]\centering
\caption{RDD-style evidence at the county-size cutoff}
\label{fig:rdd_main}
\begin{subfigure}{0.57\textwidth}\centering
  \includegraphics[width=\linewidth]{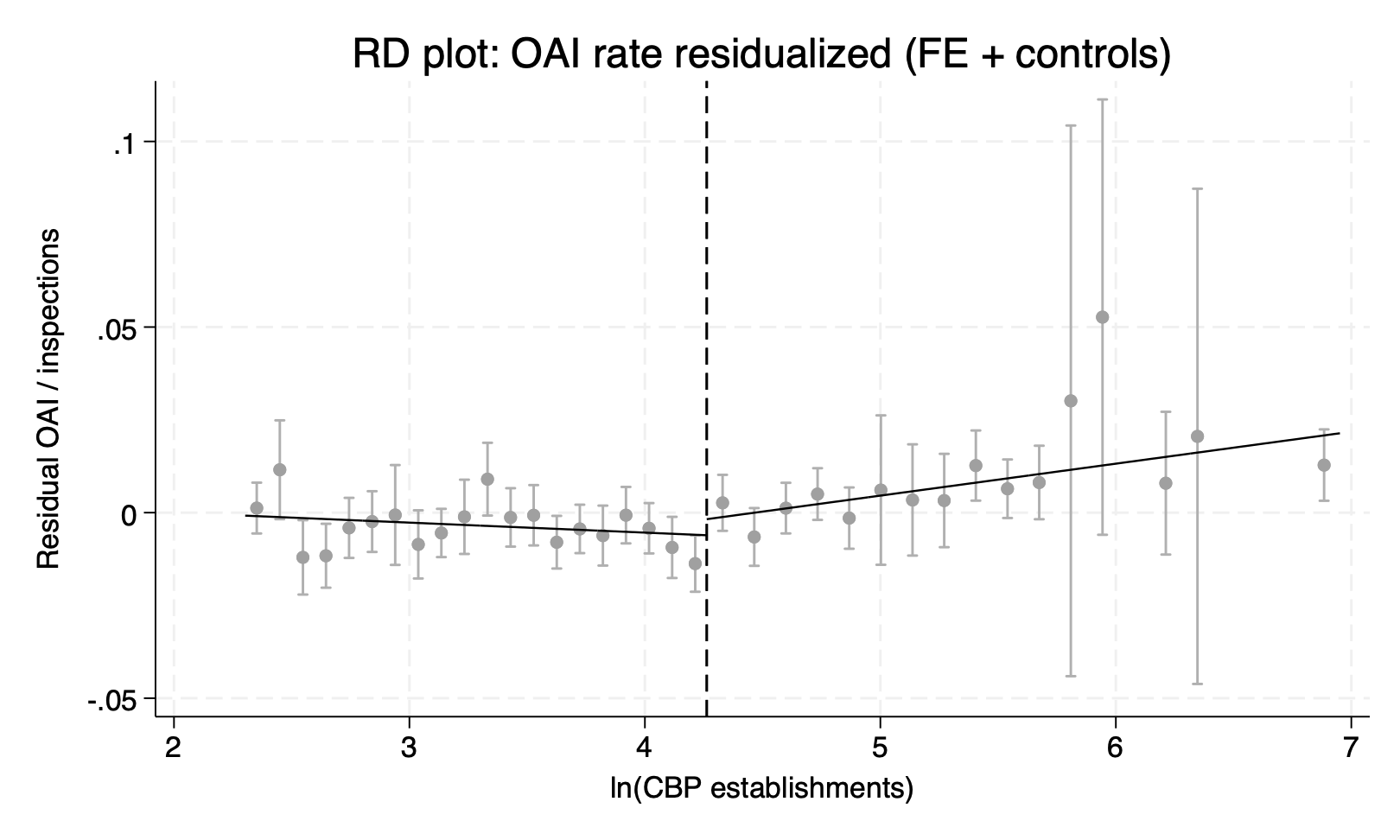}
  \caption{OAI rate}
\end{subfigure}\hfill
\begin{subfigure}{0.57\textwidth}\centering
  \includegraphics[width=\linewidth]{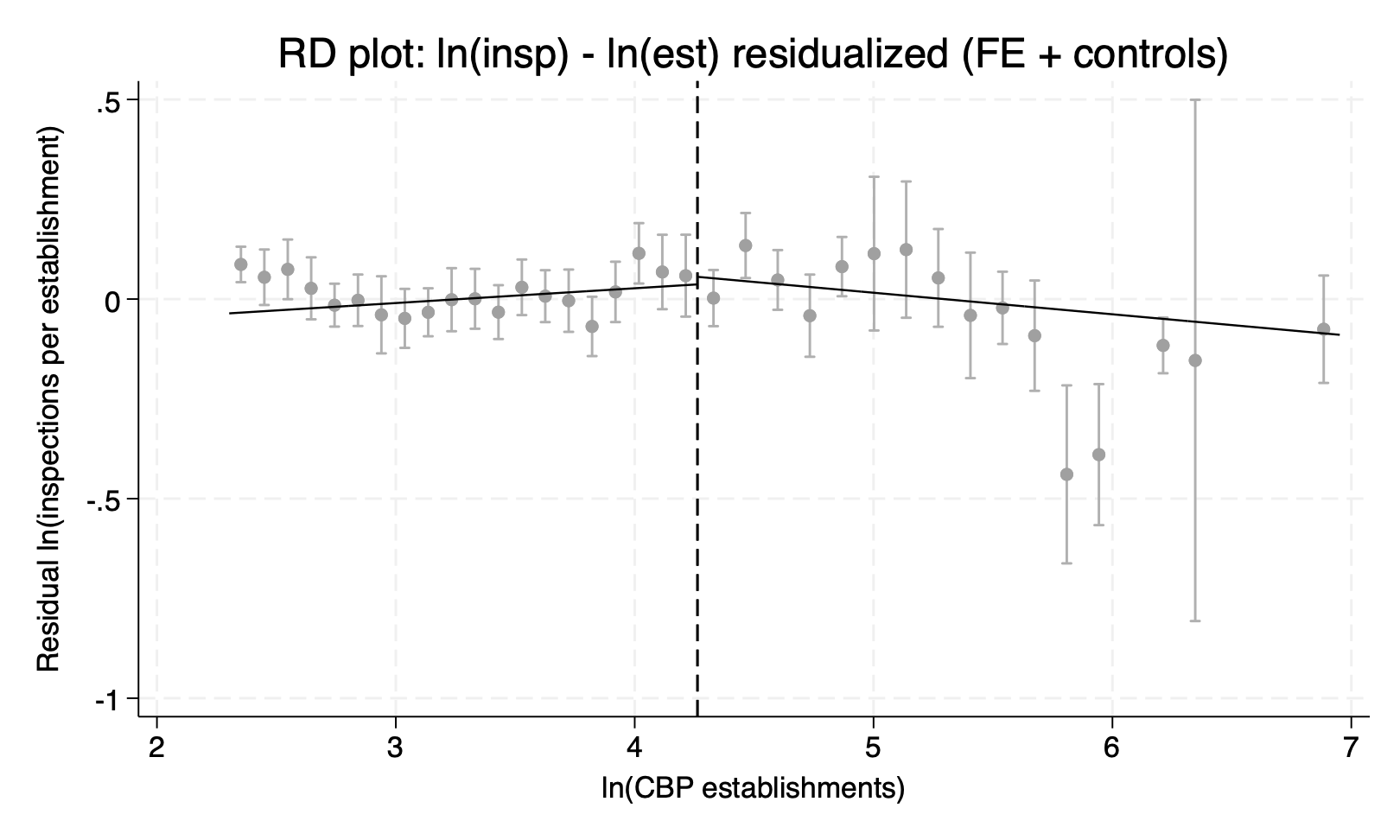}
  \caption{Monitoring effort}
\end{subfigure}\hfill
\begin{subfigure}{0.57\textwidth}\centering
  \includegraphics[width=\linewidth]{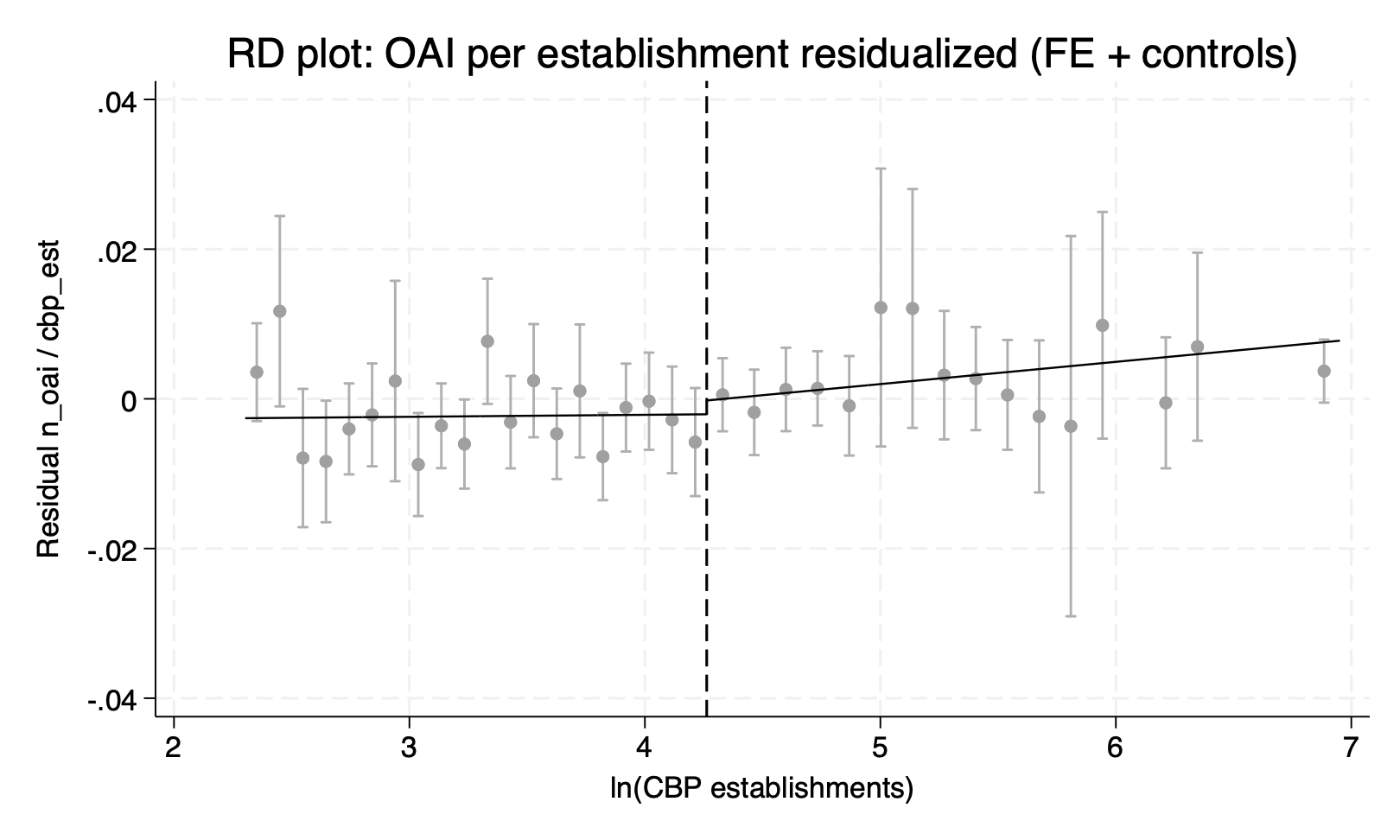}
  \caption{OAI per establishment}
\end{subfigure}

\vspace{0.6em}
\footnotesize \emph{Notes.} Running variable is $\ln(\mathrm{CBP\ establishments})$ with cutoff at $\widehat{c}=71$ (vertical line). Panels plot residuals after removing year and state fixed effects and county controls (SAIPE poverty rate, $\ln$ median household income, and LAUS unemployment). RD plots are produced using \texttt{rdplot} with a local linear fit and triangular kernel.
\end{figure}

Figure~\ref{fig:rdd_main} provides a visual check of the discontinuity-style
evidence around $\widehat c$. All panels plot residualized outcomes after
removing year and state fixed effects and county controls. The OAI-rate panel
shows a visible upward shift at the threshold, while the effort panel suggests
a flattening (or decline) in inspections per establishment above the cutoff.
These plots are not interpreted as a causal RD design---the cutoff is estimated
from the data---but they are useful for illustrating the shape implied by the
piecewise specification.

\subsection{Industry-group cutoffs: heterogeneity in where monitoring strains emerge}

We next allow the threshold to vary by industry group. For each group $g$
(NAICS-based sub-sectors), we define $S_{g,ct}=\mathrm{Establishments}_{g,ct}$ and
select a group-specific cutoff $\widehat c_g$ using the procedure in
Section~\ref{sec:design} (with the same fixed effects and controls).

Table~\ref{tab:breaks_by_group_selected_coef} shows substantial heterogeneity
in $\widehat c_g$, indicating that the scale at which monitoring outcomes begin
to deteriorate is not universal. In several groups, the OAI piecewise effects are
economically meaningful even when individual coefficients are imprecisely
estimated. The effort results are also heterogeneous: some sectors show little
change at the cutoff but a strong post-cutoff decline in effort, consistent with
monitoring becoming progressively more difficult as sector-specific local scale
increases.

\begin{table}[!htbp]
\centering
\begin{threeparttable}
\caption{Group-specific breakpoints and piecewise effects (sector-specific county--year panels)}
\label{tab:breaks_by_group_selected_coef}
\footnotesize
\setlength{\tabcolsep}{6pt}
\renewcommand{\arraystretch}{1.35}

\begin{tabular}{L{4.8cm} R{1.1cm} R{0.6cm} R{1.0cm} C{2.8cm} C{2.8cm}}
\toprule
Sector & $N$ & $c^\star$ & $\ln(c^\star)$ & OAI jump & OAI kink \\
\midrule
\multicolumn{6}{l}{\textit{Panel A. OAI piecewise effects}} \\
\addlinespace[2pt]
Bakeries and Tortilla  &
6{,}136 & 18 & 2.890 &
-0.006 (0.049) & 0.006 (0.025) \\
Other Food Manufacturing &
3{,}082 & 14 & 2.639 &
0.006 (0.079) & 0.008 (0.014) \\
Grocery and Wholesalers &
11{,}661 & 14 & 2.639 &
-0.010 (0.004) & 0.006 (0.084) \\
General Warehousing  &
6{,}474 & 14 & 2.639 &
-0.009 (0.017) & 0.006 (0.082) \\
\addlinespace[4pt]
\midrule
\multicolumn{6}{l}{\textit{Panel B. Effort piecewise effects}} \\
\addlinespace[2pt]
Sector & $N$ & $c^\star$ & $\ln(c^\star)$ & Effort jump & Effort kink \\
\midrule
Bakeries and Tortilla  &
6{,}136 & 18 & 2.890 &
0.019 (0.739) & -0.26 (0.000) \\
Other Food Manufacturing &
3{,}082 & 14 & 2.639 &
-0.108 (0.057) & -0.245 (0.0002) \\
Grocery and Wholesalers &
11{,}661 & 14 & 2.639 &
0.216 (0.000) & -0.066 (0.114) \\
General Warehousing  &
6{,}474 & 14 & 2.639 &
0.045 (0.524) & -0.205 (0.003) \\
\bottomrule
\end{tabular}

\begin{tablenotes}[flushleft]
\footnotesize
\item \textit{Notes:} The unit of observation is the \emph{county--year within the sector}. $N$ is the number of sector-specific county--year observations used for estimation after applying the sector-level inclusion criteria (e.g., minimum establishments and inspections and non-missing outcomes/controls). Because the same county--year can appear in multiple sectors, sector-specific $N$'s are \emph{not additive} and need not match (or sum to) the pooled county--year sample size in the baseline table.
$c^\star$ is selected (within each sector) by maximizing the binomial-logit likelihood for OAI outcomes with year and state fixed effects and county controls. ``Jump'' is the discontinuity at $\ln(c^\star)$ and ``kink'' is the change in slope above $\ln(c^\star)$. Entries report coefficients with two-sided $p$-values in parentheses; standard errors are clustered by county.
\end{tablenotes}

\end{threeparttable}
\end{table}

\subsection{Variation by density: how thresholds and post-cutoff slopes shift}

Finally, we examine whether thresholds vary systematically with within-group
density. Within each group $g$, we partition the sample into terciles of density
$D_{ct}$ (computed within group) and estimate a bin-specific cutoff
$\widehat c_{g,b}$ for $b\in\{1,2,3\}$.

Tables~\ref{tab:breaks_jump_bysector} and \ref{tab:breaks_kink_bysector} report the
corresponding jump and kink estimates by sector and density bin. Two features stand
out. First, the relationship between density and the \emph{level} of the estimated
cutoff is not uniform across sectors: in some cases, high-density bins are associated
with substantially larger $\widehat c_{g,b}$, while in others the cutoff shifts
little. Second, across sectors, a large share of the heterogeneity appears in the
\emph{post-cutoff slope} terms. In other words, even when the estimated threshold does
not move much, the rate at which outcomes change beyond the threshold can differ
sharply across density environments.

\begin{figure}[!htbp]\centering
\includegraphics[width=0.95\textwidth]{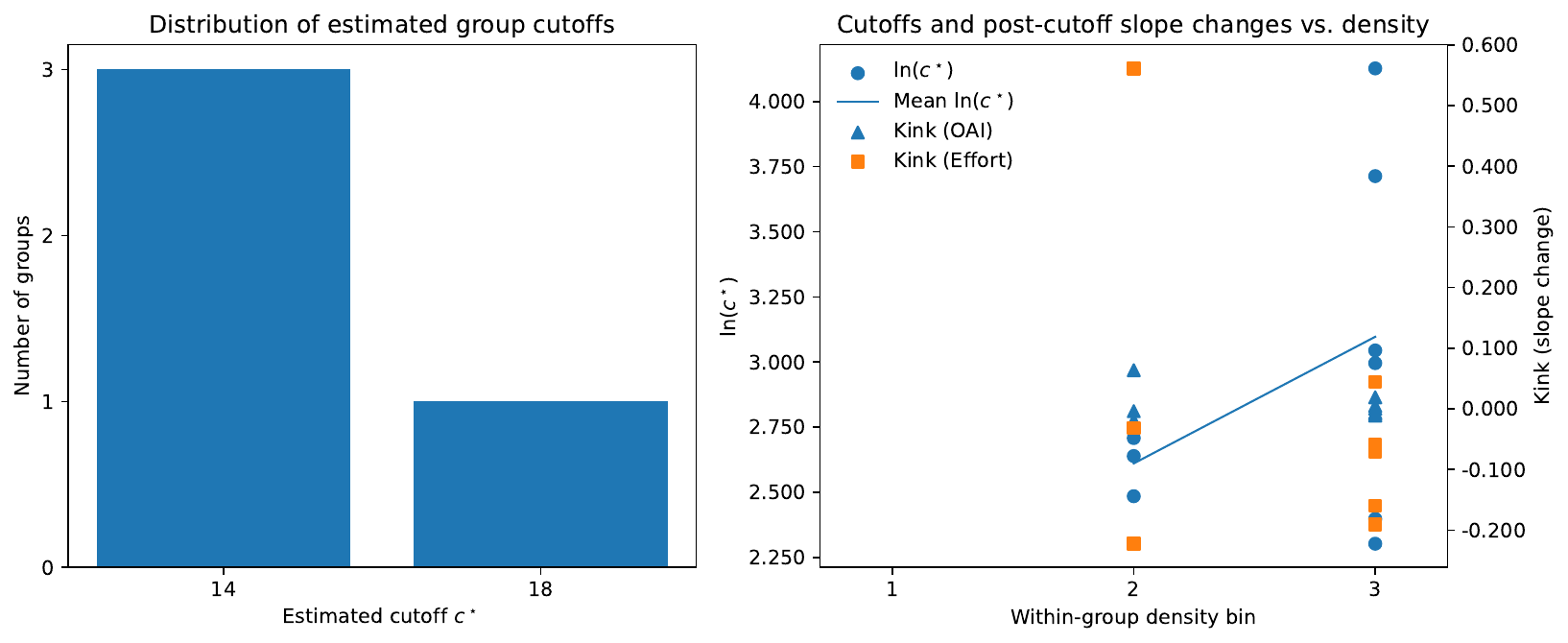}
\caption{Distribution of estimated group cutoffs (left) and how estimated cutoffs and post-cutoff slope changes vary with within-group density (right).}
\label{fig:cutoff_combined}
\end{figure}

Figure~\ref{fig:cutoff_combined} summarizes these comparisons graphically. The left
panel shows the distribution of estimated group cutoffs. The right panel plots
$\ln(\widehat c_{g,b})$ by density bin together with the corresponding post-cutoff
slope changes. The figure highlights that the empirical ``breakdown'' is not solely
about where the threshold sits in absolute terms, but also about how quickly monitoring
outcomes deteriorate once that point is reached.

\paragraph{Interpretation.}
Taken together, the national results and the heterogeneity patterns support a view
in which monitoring effectiveness depends nonlinearly on local scale. The evidence
points to two empirically distinct margins: (i) the scale at which monitoring outcomes
begin to worsen (the cutoff) and (ii) the rate at which they worsen as scale increases
further (the post-cutoff slope). The sector and density-bin results suggest that both
margins vary with the local production environment, consistent with monitoring strain
arising from organizational complexity and interconnectedness rather than a single
universal capacity threshold. Consistent with this interpretation, our auxiliary
mechanism regressions (Appendix Table~\ref{tab:triage_congestion}) show little evidence
that the breakdown is driven by systematic triage toward repeat offenders or by
mechanical delays in OAI follow-up beyond a county-year baseline cadence.

\paragraph{Robustness and placebo checks.}

Table~\ref{tab:robust_thresholds} examines sensitivity of the estimated breakpoint and
piecewise effects along several dimensions and shows that the main conclusions survive.
We vary how scale enters the model and cutoff definition (logs versus levels), and we
re-run cutoff selection using an alternative objective---minimizing the weighted RSS
from a linear-probability version of the model---in place of maximizing the grouped
binomial-logit likelihood. We also assess sensitivity to the candidate grid used in the
search: using the full set of distinct observed values recovers the baseline cutoff
($c^\star=71$), whereas a coarser percentile-step grid can select a nearby cutoff (e.g.,
$c^\star=95$), indicating that grid resolution can matter when the objective is relatively
flat over a range. Crucially, across these specifications the post-cutoff pattern is stable:
we continue to find a positive kink in OAI and a negative kink in inspection effort above
$c^\star$. To complement these exercises (see Table~\ref{tab:placebo_controls_c71} for the
baseline cutoff), we implement placebo checks using predetermined county characteristics
(poverty rate, median income, and unemployment) and apply the same piecewise specification.
These placebo regressions show no statistically meaningful discontinuity or slope change,
supporting the interpretation that the estimated breakpoint in enforcement outcomes is not
mechanically inherited from the covariates used for adjustment.

\begin{table}[!htbp]
\centering
\begin{threeparttable}
\caption{Within-group density-bin breakpoints and jump estimates (by sector)}
\label{tab:breaks_jump_bysector}
\footnotesize
\setlength{\tabcolsep}{6pt}
\renewcommand{\arraystretch}{1.15}

\begin{tabular}{L{5.0cm} C{1.0cm} R{1.0cm} R{1.0cm} R{0.8cm} R{1.1cm} C{2.6cm} C{2.6cm}}
\toprule
Sector & Bin & $N$ & $N_s$ & $c^\star$ & $\ln(c^\star)$ & $\Delta$ OAI & $\Delta$ Effort \\
\midrule
\multicolumn{8}{l}{\textit{Panel A. OAI jump estimates}} \\
\addlinespace[2pt]
Sugar and Confectionery Manufacturing            & 3 & 707    & 707    & 11 & 2.398 & 0.007 (0.188)  &  \\
Fruit and Vegetable Preserving / Specialty Foods & 3 & 653    & 653    & 10 & 2.303 & -0.009 (0.192) &  \\
Bakeries and Tortilla Manufacturing              & 2 & 1{,}947 & 1{,}947 & 12 & 2.485 & -0.000 (0.969) &  \\
Bakeries and Tortilla Manufacturing              & 3 & 1{,}955 & 1{,}955 & 41 & 3.714 & 0.009 (0.025)  &  \\
Other Food Manufacturing                         & 3 & 985    & 985    & 21 & 3.045 & 0.015 (0.001)  &  \\
Grocery and Related Product Merchant Wholesalers & 2 & 3{,}126 & 3{,}126 & 14 & 2.639 & -0.003 (0.665) &  \\
Grocery and Related Product Merchant Wholesalers & 3 & 3{,}263 & 3{,}263 & 62 & 4.127 & -0.008 (0.034) &  \\
General Warehousing and Storage                  & 2 & 1{,}961 & 1{,}961 & 15 & 2.708 & -0.019 (0.005) &  \\
General Warehousing and Storage                  & 3 & 1{,}952 & 1{,}952 & 20 & 2.996 & 0.006 (0.123)  &  \\
\addlinespace[4pt]
\midrule
\multicolumn{8}{l}{\textit{Panel B. Effort jump estimates}} \\
\addlinespace[2pt]
Sugar and Confectionery Manufacturing            & 3 & 707    & 707    & 11 & 2.398 &  & 0.167 (0.052)  \\
Fruit and Vegetable Preserving / Specialty Foods & 3 & 653    & 653    & 10 & 2.303 &  & -0.047 (0.579) \\
Bakeries and Tortilla Manufacturing              & 2 & 1{,}947 & 1{,}947 & 12 & 2.485 &  & -0.159 (0.029) \\
Bakeries and Tortilla Manufacturing              & 3 & 1{,}955 & 1{,}955 & 41 & 3.714 &  & 0.098 (0.224)  \\
Other Food Manufacturing                         & 3 & 985    & 985    & 21 & 3.045 &  & -0.030 (0.728) \\
Grocery and Related Product Merchant Wholesalers & 2 & 3{,}126 & 3{,}126 & 14 & 2.639 &  & 0.131 (0.046)  \\
Grocery and Related Product Merchant Wholesalers & 3 & 3{,}263 & 3{,}263 & 62 & 4.127 &  & 0.102 (0.089)  \\
General Warehousing and Storage                  & 2 & 1{,}961 & 1{,}961 & 15 & 2.708 &  & -0.078 (0.338) \\
General Warehousing and Storage                  & 3 & 1{,}952 & 1{,}952 & 20 & 2.996 &  & -0.008 (0.912) \\
\bottomrule
\end{tabular}
\begin{tablenotes}[flushleft]
\footnotesize
\item \textit{Notes:} Bin $1$=Low, $2$=Medium, $3$=High within-sector density. The unit of observation is the sector-specific county--year. $N$ is the number of county--year observations in the bin used for estimation; $N_s$ is the number used in the breakpoint search window (equal to $N$ here when the search and estimation windows coincide). $c^\star$ is the selected breakpoint. Entries report coefficients with two-sided $p$-values in parentheses; standard errors are clustered by county.
\end{tablenotes}
\end{threeparttable}
\end{table}

\begin{table}[!htbp]
\centering
\begin{threeparttable}
\caption{Within-group density-bin breakpoints: kink estimates (by sector)}
\label{tab:breaks_kink_bysector}
\footnotesize
\setlength{\tabcolsep}{6pt}
\renewcommand{\arraystretch}{1.15}

\begin{tabular}{L{4.9cm} C{1.0cm} R{1.0cm} R{1.0cm} R{0.8cm} R{1.1cm} C{2.6cm} C{2.6cm}}
\toprule
Sector & Bin & $N$ & $N_s$ & $c^\star$ & $\ln(c^\star)$ & Kink OAI & Kink Effort \\
\midrule
\multicolumn{8}{l}{\textit{Panel A. OAI kink}} \\
\addlinespace[2pt]
Sugar and Confectionery Manufacturing            & 3 & 707  & 707  & 11 & 2.398 & -0.011 (0.109) &  \\
Fruit and Vegetable Preserving / Specialty Foods & 3 & 653  & 653  & 10 & 2.303 & -0.008 (0.240) &  \\
Bakeries and Tortilla Manufacturing              & 2 & 1{,}947 & 1{,}947 & 12 & 2.485 & 0.064 (0.006)  &  \\
Bakeries and Tortilla Manufacturing              & 3 & 1{,}955 & 1{,}955 & 41 & 3.714 & 0.005 (0.243)  &  \\
Other Food Manufacturing                         & 3 & 985  & 985  & 21 & 3.045 & -0.006 (0.217) &  \\
Grocery and Related Product Merchant Wholesalers & 2 & 3{,}126 & 3{,}126 & 14 & 2.639 & -0.019 (0.550) &  \\
Grocery and Related Product Merchant Wholesalers & 3 & 3{,}263 & 3{,}263 & 62 & 4.127 & 0.000 (0.891)  &  \\
General Warehousing and Storage                  & 2 & 1{,}961 & 1{,}961 & 15 & 2.708 & -0.004 (0.824) &  \\
General Warehousing and Storage                  & 3 & 1{,}952 & 1{,}952 & 20 & 2.996 & 0.019 (0.003)  &  \\
\addlinespace[4pt]
\midrule
\multicolumn{8}{l}{\textit{Panel B. Effort kink}} \\
\addlinespace[2pt]
Sugar and Confectionery Manufacturing            & 3 & 707  & 707  & 11 & 2.398 &  & 0.044 (0.751) \\
Fruit and Vegetable Preserving / Specialty Foods & 3 & 653  & 653  & 10 & 2.303 &  & -0.059 (0.591) \\
Bakeries and Tortilla Manufacturing              & 2 & 1{,}947 & 1{,}947 & 12 & 2.485 &  & 0.561 (0.043) \\
Bakeries and Tortilla Manufacturing              & 3 & 1{,}955 & 1{,}955 & 41 & 3.714 &  & -0.071 (0.411) \\
Other Food Manufacturing                         & 3 & 985  & 985  & 21 & 3.045 &  & -0.069 (0.490) \\
Grocery and Related Product Merchant Wholesalers & 2 & 3{,}126 & 3{,}126 & 14 & 2.639 &  & -0.222 (0.364) \\
Grocery and Related Product Merchant Wholesalers & 3 & 3{,}263 & 3{,}263 & 62 & 4.127 &  & -0.191 (0.003) \\
General Warehousing and Storage                  & 2 & 1{,}961 & 1{,}961 & 15 & 2.708 &  & -0.031 (0.911) \\
General Warehousing and Storage                  & 3 & 1{,}952 & 1{,}952 & 20 & 2.996 &  & -0.160 (0.055) \\
\bottomrule
\end{tabular}

\begin{tablenotes}[flushleft]
\footnotesize
\item \textit{Notes:} Bin $1$=Low, $2$=Medium, $3$=High within-group density (defined within each sector). The unit of observation is the sector-specific county--year. $N$ is the number of county--year observations in the bin used for estimation; $N_s$ is the number used in the breakpoint search window (here equal to $N$ if the search window matches the estimation window). $c^\star$ is the selected breakpoint. ``Kink'' is the change in slope above $\ln(c^\star)$. Entries report coefficients with two-sided $p$-values in parentheses; standard errors are clustered by county.
\end{tablenotes}

\end{threeparttable}
\end{table}

\subsection{Mechanisms and discussion}\label{subsec:mechanisms}

Our estimates point to a nonlinear monitoring technology with a regime change at $c^{\star}$.
In the baseline specification, the OAI outcome becomes steeper above $c^{\star}$ (a positive
\emph{kink}), while monitoring effort per establishment flattens or declines (a negative
\emph{kink}). Robustness exercises show that this qualitative pattern persists across alternative
scale definitions, objective functions, candidate grids, and fixed-effect structures. In parallel,
placebo and balance checks using predetermined county characteristics (poverty, income, unemployment)
reveal no meaningful discontinuity or slope break at $c^{\star}$, and applying the cutoff search
directly to these controls does not uncover a stable ``false threshold.'' Taken together, these
findings support an interpretation in which the estimated breakpoint reflects a monitoring-relevant
regime change rather than a mechanical artifact of the sample, controls, or specification choices.

Several mechanisms could in principle generate ``too big to monitor'' dynamics. First,
\textbf{capacity constraints and convex enforcement costs} imply that detection and follow-up
resources need not scale one-for-one with the size of the regulated population. Canonical enforcement
models emphasize that monitoring is costly and enforcement resources are limited, so inspection
intensity may rise less than proportionally with scale
\citep{polinsky2000public,harrington1988enforcement}. In our setting, inspector staffing, travel
time, administrative processing, and follow-up actions may expand more slowly than county scale
$S_{ct}$, generating a post-$c^{\star}$ flattening or decline in inspections per establishment.
Second, \textbf{risk-based targeting and regulatory triage} may emerge when capacity binds: agencies
may reallocate effort toward higher-risk establishments or higher social-benefit cases, changing both
the composition of inspections and measured outcomes. Evidence from environmental regulation shows
that discretion and targeting can materially affect inspection effectiveness and compliance
\citep{duflo2018discretion}, and evidence from OSHA indicates that improved targeting can
substantially increase injuries averted per inspection \citep{johnson2023targeting}. Third,
\textbf{spillovers and connectedness (local complexity)} can generate a breakpoint even without a
hard staffing constraint. In dense production environments, compliance risks may be correlated
through shared suppliers, logistics nodes, and production networks, so violations cluster and
information from one inspection is less informative about broader compliance. If investigations and
remediation require tracing common inputs or coordinating across connected establishments, the
effective cost of monitoring can rise nonlinearly with local scale and network density
\citep{carvalho2014networks,acemoglu2015systemic}.

Our heterogeneity results naturally motivate the connectedness channel. Allowing the breakpoint to
vary by industry group yields substantially different estimated cutoffs $\widehat c_g$ across sectors
(Table~\ref{tab:breaks_by_group_selected_coef}), consistent with a mechanism governed by
\emph{sector-specific local environments} rather than a single universal county threshold. Moreover,
a large share of the economically relevant deterioration often appears in the \emph{post-cutoff
slope} rather than a discrete jump, indicating that outcomes can worsen progressively as local scale
increases within sectors. Finally, within-sector density comparisons show that density shifts both
(i) the location of the threshold and (ii) the magnitude of post-cutoff slope changes
(Tables~\ref{tab:breaks_jump_bysector} and \ref{tab:breaks_kink_bysector};
Figure~\ref{fig:cutoff_combined}), consistent with monitoring strain that depends on local
organizational complexity and interdependence.

\subsection{Why connectedness? Discriminating evidence}\label{subsec:why_connectedness}

The baseline and heterogeneity results establish a regime change in monitoring outcomes as local
scale grows. A key question is \emph{what} drives the post-$c^\star$ deterioration: a mechanical
capacity/queueing story in which larger counties triage or delay follow-up, or a
\emph{connectedness/local-complexity} story in which effective monitoring becomes harder because
violations and remediation tasks are interdependent in denser production environments. This section
summarizes three pieces of evidence that jointly favor connectedness as a first-order channel.

\paragraph{(i) Little evidence of triage toward repeat offenders.}
If the post-$c^\star$ breakdown were primarily a reallocation of scarce enforcement effort toward
establishments with recent serious violations, we would expect larger counties to exhibit more
\emph{repeat-offender targeting} among inspected facilities. Appendix
Table~\ref{tab:triage_congestion}, Column~(1), provides little support for this view: the county-scale
indicator (CBP establishments $\geq 71$) is close to zero and statistically indistinguishable from
zero in a regression for whether an inspected facility had an OAI in the prior five years, conditional
on year, product type, and project-area fixed effects. While this does not rule out more subtle
targeting along unobserved dimensions, it suggests that the deterioration above $c^\star$ is not
explained by a coarse shift toward inspecting establishments with recent OAI histories.

\paragraph{(ii) Little evidence of mechanical delay in OAI follow-up.}
A simple queueing interpretation predicts longer delays between an OAI and the next inspection in
larger counties if follow-up scheduling becomes congested. Columns~(2)--(4) of Appendix
Table~\ref{tab:triage_congestion} test this prediction using alternative normalizations of the
time-to-next-inspection after an OAI (including relative-gap and log-difference measures). Across
specifications, the point estimates are small relative to the substantial dispersion in follow-up
timing and are not statistically distinguishable from zero. These results point away from a
first-order story in which the post-$c^\star$ deterioration is driven by a systematic, mechanical
slowdown in follow-up scheduling in large counties.

\paragraph{(iii) Connectedness predicts \emph{earlier} breakdown in relative terms.}
In contrast, the heterogeneity patterns align naturally with connectedness. First, density-bin
estimates show that both the \emph{location} of the cutoff and the \emph{severity} of post-cutoff slope
changes vary across local environments (Tables~\ref{tab:breaks_jump_bysector} and
\ref{tab:breaks_kink_bysector}; Figure~\ref{fig:cutoff_combined}), consistent with a mechanism that
depends on more than the number of establishments. Second, locating the selected cutoffs within each
sector--bin distribution of establishments makes the density pattern particularly transparent.
Appendix Table~\ref{tab:cutoff_percentiles} reports the percentile position of each selected
$c^\star_{g,b}$ within the empirical distribution of $S_{g,ct}$ for that sector--bin. In the sectors
where both medium- and high-density bins are observed, the cutoff occurs \emph{earlier} in denser
counties: the percentile position of $c^\star$ falls from $89.7$ to $75.6$ in bakeries/tortilla, from $79.5$ to $63.1$ in grocery/wholesalers, and from $88.7$ to $40.4$ in warehousing. Thus, even when absolute cutoffs are higher in dense environments, the regime change
arrives at a lower \emph{rank} in the local scale distribution, consistent with effective monitoring
capacity binding sooner where establishments are more concentrated and plausibly more interconnected.

These patterns are difficult to reconcile with a purely mechanical triage or
follow-up-delay story: we find little evidence that larger counties systematically (i) concentrate
inspections on recent OAI histories or (ii) delay follow-ups after an OAI relative to baseline
inspection cadence. Instead, the evidence is consistent with a decline in \emph{effective} monitoring
in more connected environments. When local production and distribution are more intertwined (through
shared suppliers, logistics nodes, or concentrated input markets), compliance risks may be more
correlated and remediation may require coordinated follow-up across related establishments. Under
this view, scale interacts with connectedness: the same nominal inspection effort yields weaker
effective coverage once local density and interdependence cross a threshold, generating an earlier
and steeper breakdown in monitoring performance above $c^\star$.

\paragraph{Implications.}
Interpreting the evidence through a connectedness lens suggests that designing effective oversight
requires more than scaling inspection counts with the number of establishments. When production is
dense and interdependent, effective monitoring may require additional coordination capacity, better
risk signals that account for correlated exposures, and follow-up strategies designed for clustered
violations. More generally, the results emphasize that the relevant constraint is \emph{effective}
coverage: environments can become ``too big to monitor locally'' not only because staffing is finite
but because interdependence increases the complexity of diagnosing and remediating noncompliance.

\section{Conclusion}\label{sec:conclusion}

This paper documents a nonlinear relationship between local scale and FDA monitoring outcomes. Using a breakpoint-selection procedure in county--year panels, we find a regime shift at which inspection \emph{effort per establishment} flattens or declines, while the \emph{OAI outcome} becomes more sensitive to scale. This pattern is robust across alternative scale definitions, objective functions, candidate grids, and fixed-effect structures. Placebo and balance checks using predetermined county characteristics (poverty, income, unemployment) show no comparable ``false threshold,'' supporting the interpretation that the estimated break reflects a change in monitoring performance rather than covariate-driven discontinuities.

The results also emphasize that the breakdown is not characterized by a single universal cutoff. Estimated breakpoints vary substantially across industry groups and shift with within-group density, and in many cases the economically relevant deterioration appears in \emph{post-cutoff slope changes} rather than discrete jumps. These heterogeneity patterns point to a ``too big to monitor locally'' mechanism in which scale interacts with the local organization of production: monitoring becomes harder not only because the regulated population is larger, but because oversight tasks become more interdependent in denser and more connected environments. Consistent with this view, auxiliary evidence provides little support for simple triage or queueing-delay explanations, and instead suggests that effective monitoring may deteriorate through correlated noncompliance and more complex remediation once local scale and connectedness are sufficiently high.

Beyond the FDA context, the empirical framework developed here can be applied to other regulatory settings to diagnose where oversight begins to break down and whether the deterioration operates through diminished inspection intensity, reduced effectiveness conditional on inspection, or both. A central implication is that monitoring systems should be designed with an explicit view of \emph{local scale and network complexity}: policies that expand nominal inspection activity may be insufficient if the binding constraint is the complexity of follow-up and the correlated nature of risks in dense production systems.

Two limitations are worth noting. First, the inspection data reflect \emph{selected monitoring} rather than a census of establishments: FDA inspection assignments are risk-based and evolve over time, so county--year outcomes summarize performance on the inspected subset and may not map one-for-one into population-wide noncompliance. Our fixed effects, controls, and placebo checks mitigate concerns that the breakpoint is driven by broad county trends, but selection into inspection remains a constraint on interpretation.

Second, the breakpoint is \emph{estimated} from the data and is therefore subject to cutoff uncertainty. While the likelihood-profile set indicates a relatively tight range for the national cutoff and the qualitative post-cutoff pattern is robust across objectives and grids, borderline candidates can yield nearby cutoffs when the objective is locally flat. Relatedly, group- and density-specific cutoffs are estimated on smaller subsamples and may be more sensitive to sampling variation.

These limitations suggest several directions for future research. A natural next step is to strengthen identification of the congestion versus connectedness mechanisms by incorporating richer measures of local production networks (e.g., supply-chain linkages, shipment flows, or buyer--supplier concentration) and by testing whether breakpoints align with shocks to oversight capacity (budgeting, staffing, or travel constraints). Methodologically, extensions that allow for multiple breakpoints or smoothly varying transition regimes could better capture gradual congestion dynamics. Finally, applying the framework to other enforcement settings---and to outcomes that directly measure downstream harms---would help assess external validity and quantify welfare implications of monitoring breakdown at scale.

\newpage
\appendix

\section{Data Appendix}\label{sec:data_app}

This appendix documents the data sources, cleaning steps, merges, variable construction,
and sample restrictions used to build the county--year analysis files.

\subsection{Primary sources}

We combine four inputs:
\begin{enumerate}
  \item \textbf{FDA inspection microdata.} Facility-level inspection records that include inspection outcome classifications (NAI/VAI/OAI), inspection year, and facility identifiers (e.g., FEI) and location fields used for county assignment.\footnote{In the current draft, we treat the inspection classification file as the canonical source of inspection outcomes.}
  \item \textbf{County Business Patterns (CBP).} Annual county-by-industry establishment counts used to measure local regulated scale $S_{ct}$ and sector-specific scale $S_{g,ct}$.
  \item \textbf{SAIPE.} County-year poverty and income measures (poverty rate and median household income), used as county controls.
  \item \textbf{LAUS.} County-year unemployment rate, used as a county control.
\end{enumerate}

\subsection{County mapping and aggregation}

The unit of observation is a county $c$ in year $t$. We map inspections to counties using
the county FIPS code. When the FDA location information is recorded at the ZIP level,
we use a ZIP-to-county crosswalk and assign each facility to a county based on the
crosswalk mapping. We standardize the county identifier as a 5-digit string
(\texttt{countyfips}) and construct \texttt{statefips} as the first two digits.

We aggregate the inspection microdata to the county--year level as follows. Let
$Inspections_{ct}$ denote the number of inspections and $\OAI_{ct}$ the number of
inspections classified as OAI:
\[
Inspections_{ct} \equiv \sum_{i \in (c,t)} 1\{ \text{inspection}\}, \qquad
\OAI_{ct} \equiv \sum_{i \in (c,t)} 1\{ \text{classification}=\text{OAI}\}.
\]
We define the OAI rate as $\OAIRate_{ct} \equiv \OAI_{ct}/Inspections_{ct}$.

To measure the number of distinct inspected facilities, we use the FEI identifier when
available. Specifically, we tag the first observation for each facility within a
county--year and sum these tags to obtain \texttt{n\_facilities}.

\subsection{Scale and effort measures}

We merge CBP establishment counts to the county--year panel and define the baseline
scale measure as:
\[
S_{ct} \equiv \texttt{cbp\_est}_{ct},
\]
the number of relevant establishments in county $c$ and year $t$. The running variable in
the baseline specification is $\ln(S_{ct})$, stored as \texttt{ln\_est}.

We construct monitoring effort measures from inspection counts and establishments:
\[
\text{Inspections per establishment}_{ct} \equiv Inspections_{ct}/S_{ct}, \] 
\[\ln(\text{Inspections per establishment})_{ct} \equiv \ln(Inspections_{ct})-\ln(S_{ct}),
\]
implemented as \texttt{insp\_per\_est} and \texttt{ln\_insp\_per\_est2}, respectively.

For heterogeneity analyses by industry group $g$, we construct group-specific scale
$S_{g,ct}$ from CBP by summing establishment counts within the group’s NAICS
classification and repeat the aggregation and merging steps within each group.

\subsection{County controls}

We merge county-year controls from SAIPE and LAUS:
\begin{itemize}
  \item \texttt{saipe\_pov\_rate}: poverty rate,
  \item \texttt{ln\_mhi}: log median household income, computed as $\ln(\texttt{saipe\_mhi})$ when \texttt{saipe\_mhi}$>0$,
  \item \texttt{laus\_urate}: unemployment rate.
\end{itemize}
Throughout, we use a common estimation sample defined by non-missing values for the
included controls.

\subsection{Sample restrictions}

The baseline county--year panel applies the following restrictions:
\begin{enumerate}
  \item Year range restricted to the analysis window (e.g., 2009--2019).
  \item Positive scale and inspections: \texttt{cbp\_est}$>0$ and \texttt{n\_inspections}$>0$.
  \item Minimum support: \texttt{cbp\_est}$\ge$ \texttt{min\_est} and \texttt{n\_inspections}$\ge$ \texttt{min\_insp}.
  \item Non-missing county controls (for specifications that include controls).
\end{enumerate}
We cluster standard errors by county in all regressions.

\subsection{NAICS group mapping}

 Table~\ref{tab:producttype_naics_xwalk}
reports the mapping from group labels to NAICS codes used to construct $S_{g,ct}$.

\begin{table}[htbp]\centering
\caption{Proposed correspondence between FDA inspection program type and NAICS-based industry groups}
\label{tab:producttype_naics_xwalk}
\begin{tabular}{p{2.8cm} p{3.2cm} p{8.2cm}}
\hline\hline
\textbf{Inspection producttype} & \textbf{Primary FDA center} & \textbf{Typical NAICS correspondence / relation to our \texttt{food\_grp}} \\
\hline
Food/Cosmetics &
CFSAN &
\textbf{Direct match to our NAICS grouping.} Maps to food manufacturing and related distribution/storage:
\texttt{3111--3119} (excluding \texttt{3116xx} meat/poultry), plus \texttt{312111--312113} (soft drinks, bottled water, ice), and (broad definition) \texttt{4244} food wholesalers and \texttt{493110/493120} warehousing/refrigerated storage. Cosmetics are generally \emph{not} in our \texttt{food\_grp} and would require expanding the NAICS taxonomy. \\
\hline
Veterinary &
CVM &
\textbf{Partial overlap.} If the veterinary inspections are for animal food/feed, they plausibly map to \texttt{3111} (animal food) and related distribution/storage (\texttt{4244}, \texttt{4931xx}) under a broad definition. Veterinary drugs would instead align with pharmaceutical NAICS (see Drugs). \\
\hline
Drugs &
CDER &
\textbf{Outside our food taxonomy.} Typically pharmaceutical manufacturing (e.g., \texttt{3254xx} classes such as pharmaceutical preparations / medicines). No direct correspondence to \texttt{311x/312111--3/4244/4931xx}. \\
\hline
Biologics &
CBER &
\textbf{Outside our food taxonomy.} Typically biological products and related manufacturing (often overlapping with \texttt{3254xx}-type industries), not food NAICS. \\
\hline
Devices &
CDRH &
\textbf{Outside our food taxonomy.} Typically medical device manufacturing (e.g., \texttt{3391xx}, and some electronics such as \texttt{3345xx}), not food NAICS. \\
\hline
Tobacco &
CTP &
\textbf{Outside our food taxonomy.} Typically tobacco manufacturing (e.g., \texttt{31223x}) and related distribution/retail segments; not part of \texttt{food\_grp}. \\
\hline\hline
\multicolumn{3}{p{15.2cm}}{\footnotesize \textit{Notes:} This table provides a practical correspondence for organizing heterogeneity analyses. A one-to-one mapping at the facility level requires an establishment crosswalk (e.g., FEI $\rightarrow$ industry/NAICS from registration or external business directories). FDA center scopes referenced from FDA descriptions of CFSAN, CDER, CVM/animal food, and biologics.} \\
\end{tabular}
\end{table}

\section{Estimation Appendix}\label{sec:est_app}

This appendix provides additional detail on the breakpoint selection procedure,
candidate grid construction, objective functions, and inference.

\subsection{Piecewise specification}

Let $S_{ct}$ denote the county-year scale measure and let $x_{ct}=\ln(S_{ct})$ denote the
baseline running variable. For a candidate breakpoint $c$ (in levels), define:
\[
\text{post}_{ct}(c) \equiv 1\{S_{ct}>c\}, \qquad
\text{after}_{ct}(c) \equiv \text{post}_{ct}(c)\cdot \big(x_{ct}-\ln(c)\big).
\]
The estimating equation is the piecewise linear model:
\[
y_{ct} = \beta_1 x_{ct} + \beta_2 \text{post}_{ct}(c) + \beta_3 \text{after}_{ct}(c)
+ \alpha_t + \gamma_s + X_{ct}'\delta + \varepsilon_{ct},
\]
where $\alpha_t$ and $\gamma_s$ are year and state fixed effects, and $X_{ct}$ denotes
county controls (SAIPE/LAUS). ``Jump'' refers to $\beta_2$ and ``kink'' refers to $\beta_3$.

\subsection{Search window and candidate cutoff set}

We restrict breakpoint selection to a search window (e.g., $t \le \texttt{searchend}$) and
candidate cutoffs within the empirical support of $S_{ct}$:
\[
c \in [Q_{p_{lo}}(S),\, Q_{p_{hi}}(S)],
\]
where $Q_{p}(S)$ denotes the $p$th percentile of $S_{ct}$ in the search sample. We require
minimum support on each side of the candidate cutoff:
\[
\#\{(c,t): S_{ct}\le c\} \ge \texttt{min\_side}, \qquad
\#\{(c,t): S_{ct}> c\} \ge \texttt{min\_side}.
\]

In the baseline implementation, the candidate set is the set of \emph{distinct observed}
values of $S_{ct}$ within the percentile window. Some robustness checks instead use a
coarser percentile grid (e.g., every $p\_step$ percentile), which can shift the selected
cutoff when the objective function is relatively flat over a range of nearby candidates.

\subsection{Objective functions}

The baseline breakpoint is chosen to maximize the binomial-logit log-likelihood for OAI
outcomes:
\[
\widehat c \in \arg\max_{c} \ \mathcal{L}(c),
\]
where $\mathcal{L}(c)$ is the log-likelihood from a binomial GLM for $\OAI_{ct}$ out of
$Inspections_{ct}$ with the piecewise terms and the same fixed effects and controls as in
the main specification.

As a robustness check, we also consider an alternative objective based on the linear
probability model (LPM) fit to the OAI rate:
\[
\widehat c \in \arg\min_{c} \ \mathrm{RSS}(c),
\]
where $\mathrm{RSS}(c)$ is the residual sum of squares from the weighted LPM version of the
piecewise model (with weights equal to the number of inspections).

\subsection{Estimation after selection and inference}

After selecting $\widehat c$, we estimate the piecewise effects using the full analysis
sample with the same fixed effects and controls. For the OAI outcome, we estimate the
piecewise model using the OAI rate as the dependent variable with analytic weights equal
to the number of inspections. For effort outcomes, we use analytic weights based on
establishments (or group-specific exposure) to align with the interpretation of
inspections per establishment.

All standard errors are clustered at the county level. Reported $p$-values are two-sided
tests based on the $t$ distribution with residual degrees of freedom from the weighted
regression.

\subsection{Placebo and ``no other cutoff'' checks}

To assess whether the algorithm spuriously identifies thresholds, we apply the same
piecewise specification to predetermined county controls (poverty, income, unemployment)
as outcomes. We report the estimated jump and kink at the baseline cutoff and (in
supplementary output) examine the objective function across candidate $c$ when these
controls are outcomes. The placebo regressions yield no statistically meaningful jump or
kink terms and do not produce a stable alternative cutoff, consistent with the view that
the main results are not mechanical artifacts of the search procedure.

\section{Additional Tables}

\begin{table}[!htbp]\centering
\caption{Robustness: breakpoint selection and piecewise effects (county--year, 2009--2019)}
\label{tab:robust_thresholds}
\footnotesize
\begin{threeparttable}
\setlength{\tabcolsep}{5pt}
\renewcommand{\arraystretch}{1.15}

\begin{tabular}{lcccc cc}
\toprule
Spec & $c^{\star}$ & $\ln(c^{\star})$ & $N_{cy}$ & $N_{\mathrm{insp}}$ & Jump & Kink \\
\midrule

\multicolumn{7}{l}{\textit{Panel A. OAI (outcome: OAI rate)}}\\
\addlinespace[2pt]
Baseline (log, LL)                 & 71  & 4.263 & 4,027 & 141,982 & 0.006 (0.067) & 0.010 (0.001) \\
Scale in levels (LL)               & 111 & 4.710 & 4,027 & 141,982 & 0.009 (0.040) & -0.000 (0.283) \\
Alt objective (RSS)                & 71  & 4.263 & 4,027 & 141,982 & 0.006 (0.067) & 0.010 (0.001) \\
Percentile-grid candidates (p=1)   & 95  & 4.554 & 4,027 & 141,982 & 0.008 (0.029) & 0.007 (0.028) \\
Looser support (minside=80)        & 71  & 4.263 & 4,027 & 141,982 & 0.006 (0.067) & 0.010 (0.001) \\
Alt FE (region$\times$year)        & 70  & 4.248 & 4,027 & 141,982 & 0.006 (0.060) & 0.010 (0.001) \\
Alt FE (state trends)              & 73  & 4.290 & 4,027 & 141,982 & 0.006 (0.058) & 0.010 (0.001) \\

\addlinespace[4pt]\midrule
\multicolumn{7}{l}{\textit{Panel B. Effort (outcome: $\ln(\mathrm{insp})-\ln(\mathrm{est})$)}}\\
\addlinespace[2pt]
Baseline (log, LL)                 & 71  & 4.263 & 4,027 & 141,982 & 0.054 (0.373) & -0.103 (0.090) \\
Scale in levels (LL)               & 111 & 4.710 & 4,027 & 141,982 & -0.049 (0.502) & -0.000 (0.576) \\
Alt objective (RSS)                & 71  & 4.263 & 4,027 & 141,982 & 0.054 (0.373) & -0.103 (0.090) \\
Percentile-grid candidates (p=1)   & 95  & 4.554 & 4,027 & 141,982 & 0.030 (0.690) & -0.126 (0.060) \\
Looser support (minside=80)        & 71  & 4.263 & 4,027 & 141,982 & 0.054 (0.373) & -0.103 (0.090) \\
Alt FE (region$\times$year)        & 70  & 4.248 & 4,027 & 141,982 & 0.046 (0.456) & -0.104 (0.088) \\
Alt FE (state trends)              & 73  & 4.290 & 4,027 & 141,982 & 0.051 (0.428) & -0.112 (0.070) \\

\bottomrule
\end{tabular}

\begin{tablenotes}[flushleft]\footnotesize
\item \textit{Notes:} Unit of observation is county--year. All specifications include baseline county controls
(saipe\_pov\_rate, ln\_mhi, laus\_urate) and year and state fixed effects, except where alternative FE are indicated.
$N_{cy}$ is the number of county--year observations; $N_{\mathrm{insp}}$ is total inspections summed across county--years.
Entries report coefficients with two-sided $p$-values in parentheses; standard errors are clustered by county.
“Jump” is the discontinuity at $c^{\star}$; “Kink” is the change in slope above $c^{\star}$.
The “Percentile-grid candidates” row restricts the cutoff search to percentile-generated candidate values, which can shift the selected $c^{\star}$ (e.g., 95 vs 71) without materially changing the estimated effects.
\end{tablenotes}
\end{threeparttable}
\end{table}
\begin{table}[!htbp]\centering
\caption{Placebo/balance checks: county controls at $c^{\star}=71$}
\label{tab:placebo_controls_c71}
\footnotesize
\begin{threeparttable}
\setlength{\tabcolsep}{5pt}
\renewcommand{\arraystretch}{1.15}
\begin{tabular}{lccccc}
\toprule
Outcome & $N_{cy}$ & Jump & $p$-value & Kink & $p$-value \\
\midrule
SAIPE poverty rate (\texttt{saipe\_pov\_rate}) & 4,027 & 0.559 & 0.581 & -0.687 & 0.539 \\
$\ln$ median HH income (\texttt{ln\_mhi})      & 4,027 & -0.024 & 0.582 & -0.012 & 0.821 \\
Unemployment rate (\texttt{laus\_urate})      & 4,027 & 0.065 & 0.834 & -0.177 & 0.633 \\
\bottomrule
\end{tabular}
\begin{tablenotes}[flushleft]\footnotesize
\item \textit{Notes:} Unit is county--year (2009--2019). Each row reports the discontinuity (“Jump”) and slope change (“Kink”) from the piecewise specification
$y=\beta_1\ln(\text{est})+\beta_2\mathbf{1}\{\text{est}>c^\star\}+\beta_3\max\{0,\ln(\text{est})-\ln(c^\star)\}+\text{state FE}+\text{year FE}+\varepsilon$,
with standard errors clustered by county.
\item The cutoff is fixed at $c^{\star}=71$ (from the main analysis). When scanning candidate cutoffs for these control outcomes, we do \emph{not} find a distinct/meaningful breakpoint (the objective is essentially flat and no unique minimizer emerges). Consistent with this, the estimated jumps/kinks at $c^\star=71$ are not statistically significant.
\end{tablenotes}
\end{threeparttable}
\end{table}

\begin{table}[!htbp]\centering
\caption{Triage vs. Congestion Mechanisms in FDA Inspections (2009--2019)}
\label{tab:triage_congestion}
\footnotesize
\def\sym#1{\ifmmode^{#1}\else\(^{#1}\)\fi}

\begin{tabular}{l*{5}{c}}
\toprule
                &\multicolumn{1}{c}{(1)}&\multicolumn{1}{c}{(2)}&\multicolumn{1}{c}{(3)}&\multicolumn{1}{c}{(4)}&\multicolumn{1}{c}{(5)}\\
                &\multicolumn{1}{c}{Triage: Repeat OAI}&\multicolumn{1}{c}{Cong.: Rel. gap}&\multicolumn{1}{c}{Cong.: Log diff.}&\multicolumn{1}{c}{Cong.: Log diff. +1}&\multicolumn{1}{c}{Persis.: Next OAI}\\
\midrule
$N_{est}$ $\geq 71$
                &  -0.0029         &  10.2950         &   0.0457         &  -0.2709         &   0.0260  \\
                & (0.0088)         &(26.1775)         & (0.0786)         & (0.2756)         & (0.0149)         \\
\addlinespace
Year FE         &      Yes         &      Yes         &      Yes         &      Yes         &      Yes         \\
Product FE      &      Yes         &      Yes         &      Yes         &      Yes         &      Yes         \\
Project area FE &      Yes         &      Yes         &      Yes         &      Yes         &      Yes         \\
\midrule
Observations    &  235,802         &    6,976         &    4,335         &    6,976         &    7,688         \\
R-squared       &    0.017         &    0.072         &    0.103         &    0.130         &    0.021         \\
\bottomrule
\multicolumn{6}{p{0.99\textwidth}}{\footnotesize\emph{Notes:} County-clustered standard errors in parentheses.
The regressor is an indicator for counties with $N_{est}$=CBP establishments $\geq 71$.
Column (1) outcome (\emph{Repeat OAI}) equals 1 if the facility had an OAI within the prior 5 years at the time of the inspection.
Columns (2)--(4) restrict the sample to OAI inspections with a subsequent inspection observed; the baseline gap is the county-year median inter-inspection time among all inspections, so the relative gap is the OAI follow-up delay minus that county-year median.
Column (5) restricts to repeat-offender inspections with the next inspection observed; the outcome equals 1 if the next inspection is classified as OAI.
All models include year, product type, and project area fixed effects.}\\
\end{tabular}
\end{table}

\begin{table}[!htbp]\centering
\caption{Where selected cutoffs fall in the within-sector scale distribution}
\label{tab:cutoff_percentiles}
\footnotesize
\setlength{\tabcolsep}{6pt}
\renewcommand{\arraystretch}{1.15}
\begin{threeparttable}
\begin{tabular}{L{6.2cm} C{0.7cm} R{1.1cm} R{0.9cm} R{1.1cm} R{1.1cm} R{0.9cm} R{0.9cm} R{0.9cm}}
\toprule
Sector & Bin & $N_s$ & $c^{\star}$ & $\ln(c^{\star})$ & Pctl$(c^{\star})$ & Q25 & Q50 & Q75 \\
\midrule
Sugar \& Confectionery Manufacturing                    & 3 & 707   & 11 & 2.398 & 76.4 & 5  & 7  & 11 \\
Fruit \& Vegetable Preserving / Specialty Foods         & 3 & 653   & 10 & 2.303 & 73.2 & 4  & 7  & 11 \\
\addlinespace[2pt]
Bakeries \& Tortilla Manufacturing                      & 2 & 1{,}947 & 12 & 2.485 & 89.7 & 5  & 7  & 9  \\
Bakeries \& Tortilla Manufacturing                      & 3 & 1{,}955 & 41 & 3.714 & 75.5 & 15 & 24 & 41 \\
\addlinespace[2pt]
Other Food Manufacturing                                & 3 & 985   & 21 & 3.045 & 78.8 & 9  & 13 & 20 \\
\addlinespace[2pt]
Grocery \& Related Product Merchant Wholesalers         & 2 & 3{,}126 & 14 & 2.639 & 79.5 & 7  & 10 & 14 \\
Grocery \& Related Product Merchant Wholesalers         & 3 & 3{,}263 & 62 & 4.127 & 63.1 & 26 & 44 & 84 \\
\addlinespace[2pt]
General Warehousing \& Storage                          & 2 & 1{,}961 & 15 & 2.708 & 88.7 & 5  & 8  & 12 \\
General Warehousing \& Storage                          & 3 & 1{,}952 & 20 & 2.996 & 40.4 & 15 & 24 & 42 \\
\bottomrule
\end{tabular}
\begin{tablenotes}[flushleft]
\footnotesize
\item \textit{Notes:} Bin 1=Low, 2=Medium, 3=High within-sector density terciles (defined within each sector).
$N_s$ is the number of sector--bin county--years in the cutoff-search window.
Pctl$(c^{\star})$ is $100\times \Pr(S_{g,ct} \le c^{\star})$ computed within the sector--bin distribution
(in the search window). Q25/Q50/Q75 are within-bin establishment quantiles (search window).
\end{tablenotes}
\end{threeparttable}
\end{table}

\newpage

\bibliographystyle{aer} 
\bibliography{ref_cent}    

@misc{FDA_CP7303_040,
  author       = {{U.S. Food and Drug Administration}},
  title        = {Compliance Program 7303.040: State Cooperative Programs and Interstate Travel (Domestic Inspections)},
  howpublished = {Compliance Program Manual (PDF)},
  year         = {2025},
  note         = {Accessed 2026-02-03},
  url          = {https://www.fda.gov/media/131744/download}
}

@misc{FDA_LifeAfterOAI,
  author       = {{U.S. Food and Drug Administration}},
  title        = {Life after {OAI}: Understanding FDA Inspection Classifications (NAI/VAI/OAI)},
  howpublished = {Agency guidance/explainer (PDF)},
  year         = {2021},
  note         = {Accessed 2026-02-03},
  url          = {https://www.fda.gov/media/164381/download}
}

@misc{FDA_ORA_Directory,
  author       = {{U.S. Food and Drug Administration, Office of Regulatory Affairs}},
  title        = {Office of Regulatory Affairs (ORA) Directory and Field Organization},
  howpublished = {Agency directory webpage},
  year         = {2024},
  note         = {Accessed 2026-02-03},
  url          = {https://www.fda.gov/files/inspections,%20compliance,%20enforcement,%20and%20criminal%20investigations/published/ORA--Directory.pdf}
}

@misc{USDAOIG_FSIS_Inspection_2013,
  author       = {{U.S. Department of Agriculture, Office of Inspector General}},
  title        = {{FSIS}---Inspection and Enforcement Activities at Swine Slaughter Plants},
  howpublished = {Audit report (PDF)},
  year         = {2013},
  note         = {Accessed 2026-02-03},
  url          = {https://usdaoig.oversight.gov/sites/default/files/reports/2022-03/24601-0001-41.pdf}
}

@article{tchuente2025too,
  title={Too Big to Monitor? Network Scale and the Breakdown of Decentralized Monitoring},
  author={Tchuente, Guy},
  journal={arXiv preprint arXiv:2511.23320},
  year={2025}
}

@misc{CBPGeography,
  author       = {{U.S. Census Bureau}},
  title        = {County Business Patterns (CBP): Geography and Data Documentation},
  howpublished  = {Webpage},
  year         = {2024},
  note         = {Accessed February 3, 2026}
}

@article{polinsky2000public,
  title={The economic theory of public enforcement of law},
  author={Polinsky, A Mitchell and Shavell, Steven},
  journal={Journal of economic literature},
  volume={38},
  number={1},
  pages={45--76},
  year={2000},
  publisher={American Economic Association}
}

@article{harrington1988enforcement,
  title={Enforcement leverage when penalties are restricted},
  author={Harrington, Winston},
  journal={Journal of Public Economics},
  volume={37},
  number={1},
  pages={29--53},
  year={1988},
  publisher={Elsevier}
}

@article{duflo2018discretion,
  title={The value of regulatory discretion: Estimates from environmental inspections in India},
  author={Duflo, Esther and Greenstone, Michael and Pande, Rohini and Ryan, Nicholas},
  journal={Econometrica},
  volume={86},
  number={6},
  pages={2123--2160},
  year={2018},
  publisher={Wiley Online Library}
}

@article{johnson2023targeting,
  title={Improving regulatory effectiveness through better targeting: Evidence from OSHA},
  author={Johnson, Matthew S and Levine, David I and Toffel, Michael W},
  journal={American Economic Journal: Applied Economics},
  volume={15},
  number={4},
  pages={30--67},
  year={2023},
  publisher={American Economic Association 2014 Broadway, Suite 305, Nashville, TN 37203-2425}
}

@article{carvalho2014networks,
  title={From micro to macro via production networks},
  author={Carvalho, Vasco M},
  journal={Journal of Economic Perspectives},
  volume={28},
  number={4},
  pages={23--48},
  year={2014},
  publisher={American Economic Association 2014 Broadway, Suite 305, Nashville, TN 37203-2418}
}

@misc{CensusSAIPE,
  author       = {{U.S. Census Bureau}},
  title        = {Small Area Income and Poverty Estimates (SAIPE)},
  howpublished = {\url{https://www.census.gov/programs-surveys/saipe.html}},
  year={2026},
  note         = {Accessed February 6, 2026}
}

@misc{FDAInspectionClassificationDB,
  author       = {{U.S. Food and Drug Administration}},
  title        = {FDA Data Dashboard: Inspections (Inspection Classifications and Citations)},
  howpublished = {Online database and dashboard},
  year         = {2026},
  url          = {https://datadashboard.fda.gov/oii/cd/inspections.htm},
  note         = {Datasets updated weekly; accessed 2026-02-06.}
}

@article{becker1968crime,
  title   = {Crime and Punishment: An Economic Approach},
  author  = {Becker, Gary S.},
  journal = {Journal of Political Economy},
  volume  = {76},
  number  = {2},
  pages   = {169--217},
  year    = {1968},
  doi     = {10.1086/259394}
}

@article{hansen1999threshold,
  title   = {Threshold effects in non-dynamic panels: Estimation, testing, and inference},
  author  = {Hansen, Bruce E.},
  journal = {Journal of Econometrics},
  volume  = {93},
  number  = {2},
  pages   = {345--368},
  year    = {1999},
  doi     = {10.1016/S0304-4076(99)00025-1}
}

@techreport{card2015rkd,
  title       = {Inference on Causal Effects in a Generalized Regression Kink Design},
  author      = {Card, David and Lee, David S. and Pei, Zhuan and Weber, Andrea},
  institution = {IZA},
  type        = {Discussion Paper},
  number      = {9555},
  year        = {2015}
}

@article{macher2011regulator,
  title={Regulator heterogeneity and endogenous efforts to close the information asymmetry gap},
  author={Macher, Jeffrey T and Mayo, John W and Nickerson, Jack A},
  journal={The Journal of Law and Economics},
  volume={54},
  number={1},
  pages={25--54},
  year={2011},
  publisher={University of Chicago Press Chicago, IL}
}

@article{galdin2024resilience,
  title={Resilience of global supply chains and generic drug shortages},
  author={Galdin, Anais},
  journal={Princeton University manuscript},
  year={2024}
}

@article{wang2025oai,
  title={Obligatory responses to FDA inspection outcomes and future drug shortages},
  author={Wang, Yixin and Ball, George and Anand, Gopesh and Park, Hyunwoo},
  journal={Manufacturing \& Service Operations Management},
  volume={27},
  number={3},
  pages={789--807},
  year={2025},
  publisher={INFORMS}
}

@misc{FDADataDashboard,
  author       = {{U.S. Food and Drug Administration}},
  title        = {FDA Data Dashboard},
  howpublished = {Online portal},
  year         = {2026},
  url          = {https://datadashboard.fda.gov/},
  note         = {Accessed 2026-02-06.}
}

@misc{BLSLAUS,
  author       = {{U.S. Bureau of Labor Statistics}},
  title        = {Local Area Unemployment Statistics (LAUS)},
  howpublished = {\url{https://www.bls.gov/lau/}},
  year={2026},
  note         = {Accessed February 6, 2026}
}

@misc{HUDUSPSCrosswalk,
  author       = {{U.S. Department of Housing and Urban Development}},
  title        = {HUD-USPS ZIP Code Crosswalk Files},
  howpublished  = {Webpage},
  year         = {2026},
  note         = {Accessed February 3, 2026}
}

@article{acemoglu2015systemic,
  title={Systemic risk and stability in financial networks},
  author={Acemoglu, Daron and Ozdaglar, Asuman and Tahbaz-Salehi, Alireza},
  journal={American Economic Review},
  volume={105},
  number={2},
  pages={564--608},
  year={2015},
  publisher={American Economic Association 2014 Broadway, Suite 305, Nashville, TN 37203}
}

\end{document}